\documentstyle[12pt, epsf]{article}
 \textwidth
 16.4cm
 \oddsidemargin
 2.5cm
 \advance\oddsidemargin by
 -2.4cm
 \evensidemargin
 0.0cm
 \advance\evensidemargin
 by
 -1in
 \marginparwidth
 1.9cm
 \marginparsep
 0.4cm
 \marginparpush
 0.4cm
 \topmargin
 -1.5cm
 \advance\topmargin
 by
 -0.0in
 \textheight
 22.0cm
 \makeindex

 \newcommand\noi{\noindent}
 \newcommand\beq{\begin{equation}}
 \newcommand\eeq{\end{equation}}
 \newcommand\beqn{\begin{eqnarray}}
 \newcommand\eeqn{\end{eqnarray}}
 \newcommand{\la}{\langle}
 \newcommand{\ra}{\rangle}
 \newcommand{\doublespace}
 { \renewcommand{\baselinestretch}
 {1.6}
 \large\normalsize}

\begin{document}

\vspace*{3cm}

\centerline{\Large \bf Polarized Proton Nucleus
Scattering}

\vspace{.5cm}

\begin{center}
{\large B.Z.~Kopeliovich$^{1,2}$ and
T.L.~Trueman$^3$}
\\[8pt]
$^1${\sl Max-Planck Institut
f\"ur
Kernphysik,
Postfach
103980,
69029 Heidelberg,
Germany}\\

$^2${\sl Joint
Institute
for Nuclear Research, Dubna,
141980
Moscow Region,
Russia}\\

$^3${\sl Brookhaven National
Laboratory,
Upton, NY 11973,
USA}

\end{center}

\vspace{.5cm}
\begin{abstract}

We show that, to a very good approximation, the ratio of the spin-flip to the
non-flip parts of the elastic proton-nucleus amplitude is the same as for
proton-nucleon scattering at very high energy. The result is used to do a
realistic calculation of the analyzing power $A_N$ for $pC$ scattering
in the Coulomb-nuclear interference (CNI) region of momentum transfer.

\end{abstract}

\newpage

\doublespace
\noi
{\large\bf 1. Nuclear modification of the spin
amplitudes:
optical model for large nuclei}
\medskip

It is important, especially for the purposes of polarimetry, to have an
estimate of the small momentum transfer spin-flip $p A$ scattering in
terms of that for $p p$ elastic scattering \cite{bigpaper}. That is the
purpose of
this paper. We focus on the analyzing power for the proton $A_N$, which is the
asymmetry between scattering of the proton polarized up versus down with
regard to the scattering plane. Since this is known to be small, as are the
other spin-dependent amplitudes for $pp$ elastic scattering, if we work to
lowest order in this amplitude we can disregard the spin of the nucleons
within the nucleus.

We are left with only two spin amplitudes in $pp$ elastic
scattering
\beq
f(q)=f_{o}(q)
+
\vec\sigma\, \cdot
\frac{\vec k\times \vec
k'}
{|\vec k\times \vec
k'|}\
f_s(q)\
,
\label{10}
\eeq
\noi
where $\vec k$ and $\vec k'$ are the initial and
final
nucleon momenta in
c.m.,
$\vec\sigma$ are the Pauli
matrices,
$f_{o}(q)$ and $f_{s}(q)$ are the non-flip and spin-flip
amplitudes. These correspond, up to a constant normalization factor, to
the $pp$ amplitudes $\phi_+ = (\phi_1+\phi_3)/2$ and $\phi_5$,
respectively \cite{buttimore}. $\vec q=\vec k-\vec k'$ is the vector of
momentum
transfer; it is normal to $\vec k$ for small angle scattering.  The
normalization
of the amplitudes is fixed by the relation,
\beq
2\ f_{o}(0) = i\sigma^{NN}_{tot}(1 - i\rho)\
,
\label{11}
\eeq
\noi
where $\sigma_{\rm tot}$ is the $pp$ total cross section,
which is taken to be the same as the $pn$ cross section  at high
energy, and $\rho$ is the forward ratio of real to
imaginary parts of the elastic amplitudes.

We transform these to impact parameter $b$-space via

\beq
f(q) = \int d^2b \,e^{i\vec q \cdot \vec b}\, (\tilde{f}_o(b) +
i\,\vec\sigma \cdot(
\hat{ b}\times \hat{k})\,
\tilde f_s(b)).
\label{3}
\eeq
 Correspondingly, we denote the elastic proton-nucleus amplitudes by
$F_o(q)$ and $F_s(q)$; they are related by Fourier transform to the
partial amplitude $\tilde F_{o,s}(b)$ in impact-parameter representation,
just as in Eq.~(\ref{3}).

In the eikonal approximation \cite{glauber} the proton-nucleus
amplitude is expressed through the eikonal $\chi(\vec b)$ by

\beq
\tilde F(b) = i\,(1-e^{\chi(b)})\ ,
\label{30}
\eeq
where for large nuclei
\beq
\chi(\vec b)=i\,
\int d^2s\,\left[\tilde f_o(s)+i\,\vec\sigma \cdot(
\hat{s}\times \hat{k})\,
\tilde f_s(s)\right]\ T(\vec b-\vec s)\ .
\label{40}
 \eeq
 Here $\hat{k}$ and $\hat{s}$ are the unit vectors directed along $\vec
k$ and $\vec s$ respectively. $T(b)$ is the nuclear thickness function,
 \beq
T_A(b) =
\int\limits_{-\infty}^{\infty}
dz\,\rho_A(b,z)\ ,
\label{45}
\eeq
\noi
and $\rho_A(b,z)$ is the nuclear density which depends on
$b$ and and the longitudinal coordinate $z$.

If the radius of the nucleus substantially exceeds the radius of
interaction, the latter
can be neglected calculating the non-flip eikonal,
\beq
\chi_o(b)
\approx
i\,T_A(b)\, \int d^2s\, \tilde f_o(s)
= i\,T_A(b)f_o(0)\ ,
\label{50}
\eeq
which is the usual approximation.
For the spin-flip eikonal, however, such an
approximation
leads to zero result. Indeed, the ``spin-orbit
coupling''
$\vec\sigma\cdot(\vec s\times\vec k)$ leads to
cancelation
on integration over $\vec s$. The lowest order
nonvanishing
approximation
is
\beq
T_A(\vec b-\vec s)
\approx
T_A(b) -\vec
s\cdot\vec\nabla_b\,T_A(b).
\label{60}
\eeq
The corresponding eikonal is
\beqn
\chi_s(b) &= &-\frac{1}{2}\,\sigma \cdot (\hat{k}
\times \nabla_b)T_A(b) \int
d^2s \, s \,\tilde{f}_s(s) \\ \nonumber
&=& \sigma \cdot (\hat{k} \times \nabla_b)T(b)\left. \frac{df_s(q)}{dq}
\right|_{q=0} \, .
\eeqn

The spin-flip amplitude vanishes as $\sqrt{-t}$ in
the forward direction,
and so it can be represented as

\beq
f_s(q) =\mu_P(q) \,\frac{q}{2m_N} \,f_o(q) \ ,
\eeq
where,
for the strong
interaction part of the amplitude, $\mu_P(q)$ is a complex
function which is regular as $t \to 0$. Therefore
\beq
\chi(b)=\left[1
+
\frac{i \, \mu_P(0)}{2m_N}\,\vec\sigma\cdot(\hat{
k}\times\nabla_b)\right]\,i\,f_o(0)\,T_A(b) \
.
\label{90}
\eeq
From this,
the
amplitude for polarized proton - nucleus
elastic
scattering in optical approximation and the
first
order in $\mu_P$
reads,
\beq
F_o(q)=i\,\int d^2b \,e^{i\,\vec q\cdot\vec b}\,
\left(1 -
e^{i\,f_o(0)\,T_A(b)}\right)
\label{95}
\eeq
\beq
F_s(q)=\,\frac{\mu_P (0)\,f_o(0)}{m_N}\,
\int d^2b \, e^{i\,\vec q\cdot\vec
b}\,
(
\hat{q}\cdot\nabla_b)T_A(b)\,e^{i\,f_o(0)\,T_A(b)}
\label{100}
\eeq
Integrating this expression by parts
we
arrived at a surprising
result,
\beqn
F_s(q)&=& i\,\mu_P(0)\,\frac{q}{2m_N}\,
\int
d^2b\,e^{i\,\vec q\cdot\vec b}\,
\left(1-e^{if_o(0)T_A(b)}\right)\nonumber\\
&=& \mu_P(0)\,\frac{q}{2m_N}\,F_o(q)\ :
\label{110}
\eeqn
both the non-flip and the spin-flip parts of
the
elastic amplitude have the same nuclear
form factor $F_o(q)\,$!

This is not a trivial conclusion since the homogeneous
central nuclear region does not contribute to the
spin-flip amplitude, which according to Eq.(\ref{60}) is proportional to 
the
derivative
of nuclear thickness, {\it i.e.} gains its value only from the nuclear
periphery. On the contrary, the non-flip part of the
amplitude is large at $b < R_A$ and small at the nuclear edge.
This is compensated for partially by the fact that the derivative near the
nuclear surface is large, proportional to the nuclear radius $R_A$, and
partially due to the variation of the phase factor around the periphery
which yields on integration a factor $i q R_A$. Therefore, $F_s$ has the
same $A$ dependence as $F_o$, i.e. $A^{2/3}$. Furthermore, their relative
phase is independent of $q$.

This result requires that the nuclear size be much larger than the
slopes for $pp$ scattering; i.e.
$R_A^2 \gg 2B_o$ and $R_A^2 \gg 2B_s$. Since $2B_o$ is about $20\,
GeV^{-2}$ and $R_A^2$ ranges from about $50\, GeV^{-2}$ to $500\, GeV^{-2}$
from $He$ to $Pb$ \cite{deVries}, this strong inequality is not satisfied over
much of the periodic table and corrections need to be taken into account.

\noi
{\large\bf 2. Finite size corrections and a little theorem}

The presentation in previous section was maximally simplified
for the sake of clarity. Some corrections need to be discussed.
In the case of smaller nuclei,
one should use more accurate approximations.
Again, following Glauber \cite{glauber}, for a nucleus of $A$ nucleons with
wave
function in coordinate space $u(x_1, \dots x_A)$ the amplitude for $pA$ elastic
scattering is given by
\beq
F(q) = i \int d^2b\, e^{i \vec{q} \cdot \vec{b}} \left \{\int d^3x_1 \dots
d^3x_A |u(x_1, \dots x_A)|^2 \left[1-\prod_j(1 + i
\tilde{f}(\vec b-\vec s_j)\right]\right\}
\label{15}
\eeq
where $s_j$ is the transverse component of $x_j$. We assume that the $pp$
and $pn$ amplitudes are the same at the energy of interest and continue to
neglect the spin of the nucleons within the nucleus.

For a specific nucleus with known wave-function one could do this
numerically with an assumed form for the scattering amplitudes $f$. We
will do this for an interesting case shortly. Before doing so we will prove
a theorem which yields a very general result for a special case. Let us
assume that Eq.~(10) is valid with $\mu_P$ {\it independent of} $q$. This is
true in some models and likely not to be too far wrong. Then it is easy to
show that
\beq
\tilde{f}_s(b) = \frac{\mu_P}{2m_N} \frac{d\tilde{f}_o(b)}{db}.
\label{16}
\eeq
Then we have
\beqn
\frac{\mu_P}{2 m_N}\frac{\partial}{\partial b}\,\left[1-\prod_j(1 +
i \tilde{f}(\vec b-\vec s_j)\right]
&=&-i\,
\frac{\mu_P}{2 m_N}\sum_i
\frac{\partial
\tilde{f}_i(\vec b-\vec s_i)}{\partial b}\prod_{j
\neq i}\Bigl[1 + i\, \tilde{f}(\vec b-\vec s_j)\Bigr]
\nonumber \\
&=&-i\,\sum_i
(\hat{b}\cdot\widehat{b-s_i})\tilde{f}_s (b-s_i)\prod_{j
\neq i}\Bigl[1 + i\, \tilde{f}(b-s_j)\Bigr].
\label{17}
\eeqn

Now expand Eq.~(\ref{15}) keeping linear terms in $f_s$, take the trace
with $\vec \sigma\cdot\hat{b}\times\hat{k}/2$ of the factor in brackets;
this operation will produce the argument of the integral for $i
\tilde{F}_s(b)$. This is easily seen to match the structure given in
Eq.(\ref{17}). Thus
 \beq
\tilde{F}_s(b) = \frac{\mu_P}{2m_N} \frac{d\tilde{F}_o(b)}{db},
\eeq
and by the inverse of the process used to obtain Eq.(17) we obtain as a
theorem, with very weak assumptions on the nuclear structure,
\beq
F_s(q) = \frac{\mu_P}{2m_N} F_o(q).
\eeq
The result obtained in Section 1 may be seen as a special case of this
since for $R_A^2 \gg 2B_o$ and $R_A^2 \gg 2B_s$ the two amplitudes have
effectively the same shape in $q$ over the relevant range.

 Long ago Bethe \cite{bethe} carried out calculations with similar goals
to ours for low-energy, potential scattering. He cites there the analogous
result to Eq.~(\ref{110}) as
obtained by K\"{o}hler and by Levintov \cite{kohler}, demonstrating how
widely applicable this
relation is. See also \cite{bourrely}.

We test the robustness of this result by evaluating Eq.~(18) numerically 
for
$H\!e^4$, dropping the assumptions that the slopes
of the non-flip and flip amplitudes are the same and that the nucleus is
large. We adopt the independent particle wave function
\beq
u(x_1,x_2,x_3,x_4)= \prod_i u(x_i)
\eeq
with
\beq
u(x) = N e^{-\frac{x^2}{2 R_A^2}}
\eeq
for each of the 4 nucleons in the nucleus. We follow the method used by
Bassel and Wilkin \cite{bassel} many years ago for calculating the non-flip
amplitude. Their results agree rather well with data at low energy up to
about the first diffraction dip so it should be a useful indicator of the
robustness of this result. As before we calculate to first order in $f_s$.
We use the exponential forms for the scattering amplitude near $t=0$,
neglecting the real parts; this should be sufficiently accurate for the
small-$t$ predictions:
\beqn
2 f_o(q) &=& i\,\sigma_{\rm tot} e^{-\frac{q^2 R_o^2}{4}} \\ \nonumber
2 f_s(q) &=& i \, \frac{\mu_P}{2 m_N}\, \sigma_{\rm tot}\, e^{-\frac{q^2
R_s^2}{4}}.
\eeqn
Define $F_s^{\rm red}$ by $F_s(q) = q\,\mu_P/2\,m_N\,F_s^{\rm red}(q)$. We will
use the parameters $R_A^2=50\, GeV^{-2}$ \cite{deVries}, $R_o^2 =24 \,
GeV^{-2}$ \cite{bigpaper} and will vary $R_s$ from small values up to
$1.4\,R_o$. In
Fig.1 the results of this calculation are displayed by plotting
$F_s^{\rm red}(0)/F_o(0)$ as a function of $R_s/R_o$.
\begin{figure}[thb]
\centerline{\epsfbox{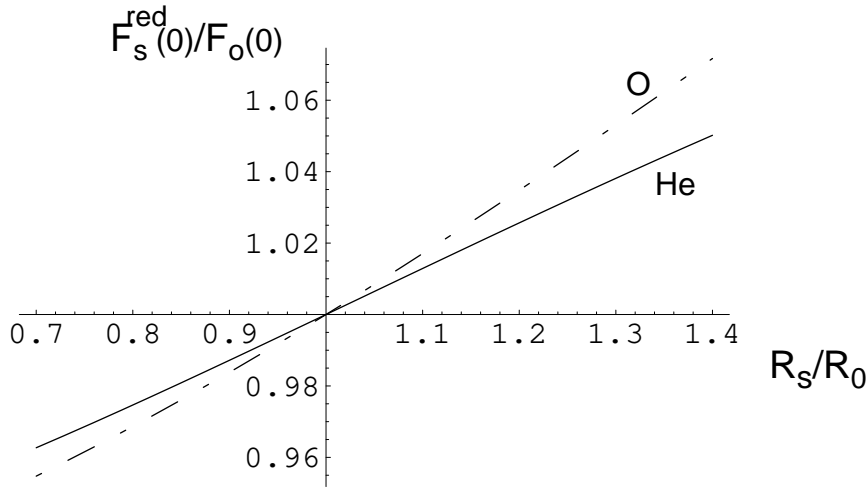}}
\medskip
\caption{\sl The ratio of the reduced nuclear spin-flip amplitude to the
non-flip amplitude at $t=0$ as a function of the ratio of the square-root
of the slopes $R_s/R_o$ for $H\!e^{4}$ and $O^{16}$.}
\end{figure}

We see that it goes through 1 at $R_s=R_o$ as required by our theorem and
that it varies by less than about $\pm 5 \% $ over this rather large range
of variation for the slopes. Thus, within the approximations of standard
multiple scattering theory we can be confident that the ratio
of flip to non-flip for $pA$ scattering will be reasonably close to that for
$pp$ scattering.

We have carried through the same calculation for $O^{16}$
using again the wave-functions of \cite{bassel}. The results are also shown
in Fig.1. The sensitivity to $R_s/R_0$ is slightly greater than for
$H\!e^4$, but still not very large. In order to estimate how this goes as the
nuclei grow we have modeled the problem with all Gaussian wave functions with
radius growing as $A^{1/3}$ and taken the derivative with respect to
$R_s/R_0$ of the amplitude at $t=0$ and $R_s/R_0=1$. This is shown as a
function of $A$ over a very wide range in Fig.2. We see that the maximum
correction occurs near oxygen, but the effect falls off very slowly with
$A$ and is still a few percent at lead.

\begin{figure}[thb]
\centerline{\epsfbox{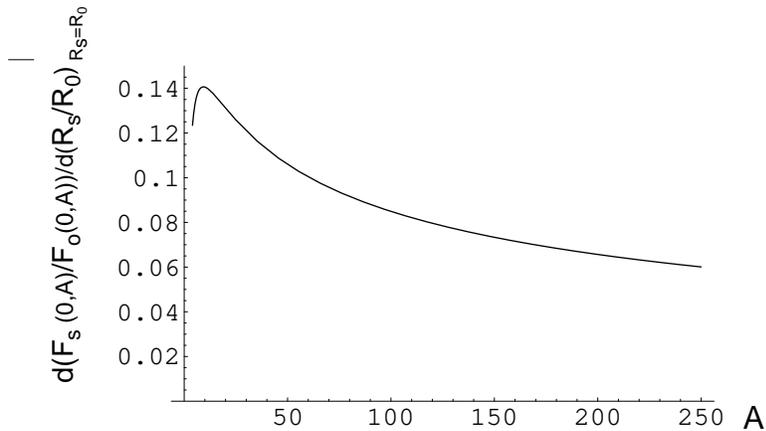}}
\medskip
\caption{\sl The derivative of the ratio of the reduced nuclear spin-flip
amplitude to the non-flip amplitude at $t=0$ and $R_s/R_0=1$ as a function of
the nuclear number A.}
\end{figure}
\newpage

\noi
{\large\bf 3. Inelastic shadowing}

A potentially serious correction to these calculations, especially at higher
energy, is inelastic shadowing. In the Glauber-type calculations only
multiple elastic scattering is taken into account, whereas diffractive
production of excited states which rescatter back into the proton state
should also be taken into account. This was emphasized by Gribov long ago
\cite{gribov}. Furthermore, there is experimental evidence in the non-flip
scattering that inelastic shadowing is important at high energy
\cite{murthy}. One does not have the knowledge of inelastic spin-flip
scattering in the hadronic basis in order to do a direct phenomenological
calculation of this correction.

An effective way to think about
these corrections in all orders is to switch to the basis of interaction
eigenstates in which the amplitude matrix is diagonal
\cite{kl1}. In QCD such a basis is given by the color-dipole light-cone
representation \cite{zkl} widely used nowadays. Thus, Eq.(5) should be
replaced by
\beq
\tilde F(b) = i\,\left\la 1-e^{\chi(b)}\right\ra\ ,
\label{3.10}
\eeq
where the eigenvalue of the amplitude is averaged over all eigenstates
(in QCD they are the Fock components with definite transverse separations).
On the contrary, in the Glauber approximation only the exponent in Eq.(5)
would be averaged ($\chi\Rightarrow\la\chi\ra$). The difference between these
two ways of averaging is just the Gribov's inelastic correction \cite{kl1,zkl}.

The eigenstate representation for inelastic corrections is rather simple
and allows
to sum up all of them in all orders. However, it needs quite high energies
to prevent the eigenstates from mixing (in the dipole picture the parton
separations should be frozen for the time of propagation through the nucleus).
At medium high energies this condition is not met and
the amount of inelastic shadowing shrinks. This
is controlled by the longitudinal nuclear form factor \cite{kk},
\beq
F^2_A(q_{\scriptscriptstyle L}) = \frac{1}{A\,\la T_A\ra}
\int d^2b\,\left|\int\limits_{-\infty}^{\infty}dz\,\rho_A(b,z)\,
e^{{\displaystyle i\,q_{\scriptscriptstyle L} z}}\right|^2\ ,
\label{3.20}
\eeq
where
\beq
\la T_A\ra = {1\over A}\int d^2b\,T_A^2(b)
\label{3.21}
\eeq
 is the mean nuclear thickness; $q_L=(M^2-m_N^2)/2E_p$, $E_p$ is the proton
lab frame energy and $M$ is the the mass of the diffractively excited proton,
which one should integrate over. Calculations and data \cite{murthy} show
that the inelastic corrections are quite small at the AGS energies and the
Glauber approximation works pretty well.

On the contrary, at the high energies of RHIC one can rely on the
opposite limit of frozen eigenstates (\ref{3.10}), or $F^2_A(q_L)=1$.
Correspondingly, this modification of the amplitude should be applied
to the non-flip spin amplitude Eq.~(\ref{95}) and to the spin-flip
amplitude Eq.~(\ref{100}) and the upper line of (\ref{110}). The result for
the spin-flip amplitude is not obvious since $\mu_P$ may
correlate to eigenstates and should be involved into the averaging.

For the lowest
Fock component of the proton, $|3q\ra$, we expect that averaging of
$\mu_P$ weighted with the wave function squared decouples from
averaging of the exponential in the upper row of Eq.~(\ref{110}).
This is certainly true if the spin-flip part originates solely from the
anomalous color magnetic moment of the quark \cite{dubna}.
However, it can also be a result of the Melosh spin rotation effect.
Although helicity is
invariant relative to a longitudinal Lorentz boost, the proton and the
quark helicities are
defined relative to different axes because of the transverse motion of the
quarks. It turns out,
however, that the spin rotation corrections cancel if the radial part of
the proton wave
function is symmetric. Only if the proton wave function is dominated by
configurations with a small diquark does the proton-Pomeron vertex acquire
a spin-flip part
$\mu_P$
\cite{kz}. It is sensitive to the smallest size in the proton (diquark), while
$f_0(0)$ in the exponent in Eq.(\ref{110}) is sensitive to the largest
inter-quark spacing in
the proton. Therefore, integrations over these two distances
factorize and the relation Eq.(\ref{110}) is preserved.

\begin{figure}[thb]
\centerline{\epsfbox{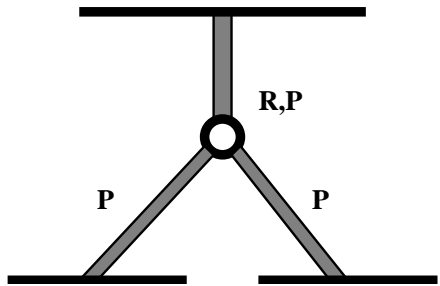}}
\medskip
{\caption[Delta]
{\sl The triple-Regge graph, where a unitarity cut of the upper
leg corresponds to the $|3q\ra$ (PPR) or $|3q\,nG\ra$ (PPP) Fock
components.}
\label{3R}}
\end{figure}

One can also look at this problem from the point of view of the
triple-Regge phenomenology and evaluate the corrections.  The Fock
state $|3q\ra$ of the proton and the higher components $|3q\,nG\ra$
in which the slowest parton is a valence quark correspond to the
triple Regge graph PPR shown in Fig.~\ref{3R}.
The leading
Reggeons $R=\omega,\ f$
are known to have a small spin-flip
amplitude, and there is a danger that they may affect the eikonal
relation Eq.(\ref{110}) via corresponding inelastic corrections.

On expanding the exponential in Eq.~(\ref{3.10}) one finds the
relative inelastic correction in the lowest order for the non-flip
amplitude \cite{kk},
\beq
-\,\frac{\Delta\tilde F(b)}{\tilde F(b)}=
{1\over 4}\,\la T_A\ra\,
\left(\frac{d\sigma_{sd}/dt}
{d\sigma_{el}/dt}\right)_{t=0}\ ,
\label{3.30}
\eeq
where $\la T_A\ra$ is the mean nuclear thickness function
defined in Eq.(\ref{3.21}).

 The forward single diffractive $pp\to pX$
cross section corresponding
to the $PPR$ contribution integrated over effective mass of
the produced system $X$ reads,
\cite{kklp}
\beq
\left.\frac{d\sigma^{PPR}_{sd}}{dt}\right|_{t=0}=
2\,G_{PPR}(0)\,\sqrt{\frac{s_0}{M_0^2}}\ ,
\label{3.50}
\eeq
 where the triple-Regge coupling $G_{PPR}(0)=2-3\,{\rm mb/GeV}^2$ was
fitted to data in \cite{kklp}; $s_0=1\,GeV^2$; $M_0\sim 1\,GeV$ is
the bottom limit for integration over masses.

Thus, the energy independent fraction of low-mass diffractive
excitation of (the PPR term) relative to the elastic
cross section is,
\beq
\left(\frac{d\sigma^{PPR}_{sd}/dt}
{d\sigma_{el}/dt}\right)_{t=0}
\approx 0.06\ .
\label{3.60}
\eeq
 The corresponding inelastic correction Eq.(\ref{3.30})  grows
$\propto A^{1/3}$ and ranges from about $1\%$ for carbon to about $
2\%$ for lead. These values demonstrate that a possible deviation
from relation Eq.(\ref{110}) caused by inelastic correction related to
the $PPR$ triple-Reggeon term, or the $|3q\ra$ Fock component of the
proton, is expected to be very small. This estimate also confirms
that the approximation of lowest order in multiple scattering
expansion in Eq.(\ref{3.30}) is rather accurate for all nuclei.

The higher order Fock components $|3q\,nG\ra$ in which the
parton carrying the least fraction of the light-cone momentum is a
gluon, correspond to triple-Pomeron term $PPP$.
We do not expect any substantial deviation from relation Eq.(\ref{110})
due to the related inelastic corrections either. Indeed,
radiation of a gluon with small fraction $x\ll1$ of the light-cone
momentum does not flip the quark helicity \cite{kst1}, the same as
the $t$-channel gluons in the Pomeron. One can reformulate this
statement in terms of the standard Regge phenomenology.

One way of doing this is to use
the fan diagrams introduced by Schwimmer \cite{schwimmer}, and
further developed recently by Bondarenko et al. \cite{bondarenko}. The
approximations used are best justified for large nuclei, and it is
not clear at present how accurate the results are quantitatively.
Nevertheless, we apply it to the problem at
hand. The typical fan diagram is shown in Fig.4.
\begin{figure}[thb]
\centerline{\epsfbox{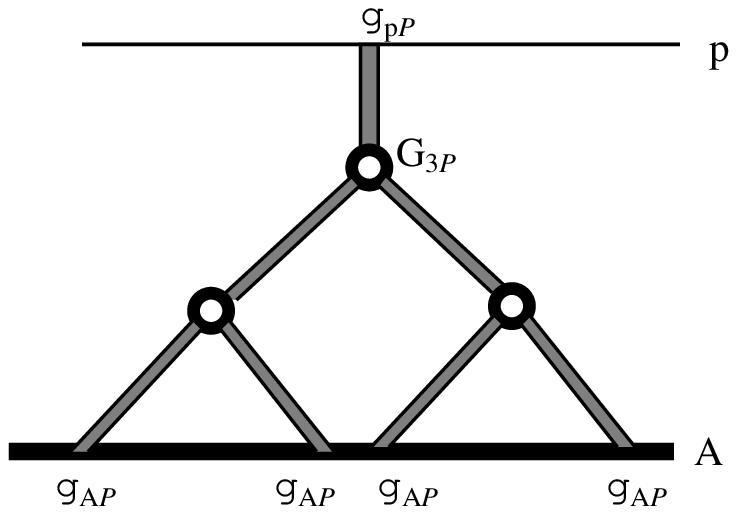}}
\medskip
\center {\caption[Delta]
\sl Typical Pomeron fan diagram. $g_{A P}(b)= A \frac{g_{{\rm p} P}}{\pi
R_A^2 s_0} \exp(-b^2/R_A^2)$}
\label{fan}
\end{figure}
The crucial feature is that
the incident proton couples to a single Pomeron, which is taken to be a pole
a small amount $\Delta$ above 1. This then branches into a fan of
Pomerons, via the triple Pomeron coupling $G_{3P}$, which then couple
to the nucleus. In order to sum these graphs, the $q^2$-dependence of the
propagators and the various vertices (except the coupling to the nucleus)
is neglected with regard to the rapid $q^2$ dependence coming from the
nucleon form factor.
The sum of these graphs in the form given by \cite{bondarenko} is
\beq
\tilde{F}^{\rm fan}_o(b) = i (s/s_0)^{\Delta}\, (A/2\pi R_A^2)\,
\frac{g_{pP}^2}{1 +
\kappa_A(b)} e^{-b^2/R_A^2},
\eeq
where
\beq
\kappa_A(b) = g_{AP}(b) \, G_{3P} \, \frac{(s/s_0)^{\Delta}
-1}{ \Delta}.
\eeq
$g_{{\rm p} P}$ denotes the Pomeron non-flip coupling to the proton which is
constant in this
approximation. By factorization of the Pomeron coupling, the spin-flip
coupling is
$g_s(q) =\mu_P\, (q/{2m_N}) g_{{\rm p} P}$. Therefore  to this approximation
the relation
$F_s(q) =
\mu_P\, (q/{2m_N}) \, F_o(q)$ is trivially maintained. Correspondingly,
\beq
\tilde{F}_s^{\rm fan}(b) = (\mu_P/{2m_N})\, \frac{d \tilde{F}_o^{\rm
fan}(b)}{db}.
\eeq
It may be that this approximation is improved by eikonalizing it in the
usual way
\cite{schwimmer, bondarenko}:
\beq
\tilde{F}^{\rm fan} \to -i\, \chi(b).
\eeq
This improved approximation, call it $\tilde{F}^{\rm eik}(q)$ then clearly
satisfies
\beqn
\tilde{F}_s^{\rm eik}(b) &=&(\mu_P/2m_N)\, \frac{d \tilde{F}_o^{\rm
fan}(b)}{db} e^{i\,\tilde{F}_o^{\rm fan}(b)} \\ \nonumber
&=& i\, (\mu_P/2m_N)\, \frac{d (1 -e^{i\,\tilde{F}_o^{\rm fan}(b)})}{db} \\
\nonumber
&=& \frac{d \tilde{F}_o^{\rm eik}(b)}{db},
\eeqn
and so the relation is preserved in this approximation as well \cite{bourrely}.
Only the inelastic corrections
related to small mass excitation, or $PPR$ triple-Regge term, cannot
be included in eikonalization, but they are found above to be small.

It is clear from these estimates that the relation Eq.~(18) is very 
robust;
however, we know from Section 2 that it cannot be exact. The inelastic
corrections important at high energies are more poorly under control and it
would be desirable to eventually have better estimates of these.

\noi
{\large\bf 4. Polarized proton-nucleus scattering and the spin-flip
Pomeron amplitude}

In this section we will use the results of the preceding sections, in
particular Eq(\ref{110}), to calculate the analyzing power $A_N$ for $pA$
scattering in the small $t$ region, $|t| \le 0.1 {\rm GeV}/c^2$. Since
$\mu_P$ is not disturbed by nuclear effects, one can use elastic $pA$
scattering to determine this parameter. Although many experimental and
theoretical results restrict
$\mu_P$ to be less than about $0.1-0.2$ \cite{bigpaper}(see also
\cite{enrico}),
none of them is strong enough or
sufficiently reliable. A promising way to fix $\mu_P$ from data is
to measure the analyzing power $A_N$ of scattering of protons with known
polarization
off protons \cite{kz,larry} or nuclei \cite{k} in the Coulomb-nucleus
interference region of $q^2$ \cite{kl}.
Hunting for the spin-flip part of the Pomeron amplitude
it is important to disentangle it from the contribution of secondary Reggeons
which is still quite important at medium high energies.
This is an important advantage of nuclear
targets which either completely eliminate
the main source of the spin-flip amplitude if the nucleus is isoscalar,
the isovector Reggeons $\rho$ and $a_2$,
or suppress them by $1/A$.

The $t$ - dependence of $A_N$ in polarized elastic $p-A$
scattering in the CNI region is similar to that in $pp$ scattering. In
particular, they both
have a distinctive peak in the neighborhood of $t=-.002 {\rm GeV}/c^2$.
However, the spin
asymmetry as function of $t=-q^2$ is substantially modified by nuclear
effects \cite{k} compared
to $pp$ scattering over the range we are interested in here
\cite{kl,buttimore,bigpaper}. In the notation of \cite{bigpaper},
\beqn
&& \frac{16\,\pi}{(\sigma^{pA}_{tot})^2}\,
\frac{d\,\sigma_{pA}}{d\,t}\,A^{pA}_N(t) =
\frac{\sqrt{-t}}{m_N}\,
F_A^h(t)\,\biggl\{F_A^{em}(t)\,\frac{t_c}{t}\,
\Bigl[(\mu_p-1)(1-\delta_{pA}\,\rho_{pA})
\nonumber\\
&-&  2\,({\rm Im}\,r^{pA}_5-\delta_{pA}\,{\rm Re}\,r^{pA}_5)\Bigr]
- 2\,F_A^h(t)\Bigl({\rm Re}\,r^{pA}_5 -
\rho_{pA}\, {\rm Im}\,r^{pA}_5\Bigr)\biggr\}\ ,
\label{4.40}
 \eeqn
where
 \beqn
\frac{16\,\pi}{(\sigma^{pA}_{tot})^2}\,
\frac{d\,\sigma_{pA}}{d\,t} &=&
\left(\frac{t_c}{t}\right)^2\,\Bigl[F_A^{em}(t)\Bigr]^2 -
2\,(\rho_{pA}+\delta_{pA})\,\frac{t_c}{t}\,
F_A^h(t)\,F_A^{em}(t)\,\nonumber\\
&+& \left(1+\rho_{pA}^2-\frac{t}{m_p^2}\,|r^{pA}_5|^2\right)\,
\Bigl[F_A^h(t)\Bigr]^2\ .
\label{4.50}
 \eeqn 
 Here $t_c=-8\pi Z\alpha/\sigma^{pA}_{tot}$;
$r_5 = \mu_P (i + \rho)/2$ and, from Eq(\ref{110}), $r_5^{pA}=
\mu_P (i + \rho_{pA})/2$. $F_A^h(t)$ and $F_A^{em}(t)$ denote the
hadronic and electromagnetic form factors, which we will calculate. (We
neglect the difference between the Dirac and Pauli form factors of the
proton which contribution to the asymmetry is negligibly small.) They
have significant $t$-dependence, as does $\rho_{pA}$ the ratio of the
real- to imaginary-part of the $pA$ non-flip amplitude, over the range of
interest here.  Finally, $\delta_{pA}$ denotes the so-called Bethe phase
\cite{bethe,cahn}; we will use a recently improved calculation \cite{kt}.
Although it is higher order in $\alpha$ it has an important effect at the
level we are calculating.

The method we use should be applicable to many nuclei (see
\cite{k,jacques}); here we will work it
out in detail for
carbon and  use
a realistic harmonic oscillator parameterization for the nuclear
density,
 \beq
\rho_C(r)=\left({a\over\pi}\right)^{\frac{3}{2}}\,
\left(2  + \frac{8}{3}\,a\,r^2\right)\, \exp{(-a\,r^2)}\ ,
\label{4.25}
 \eeq
 where $a=0.0143\,GeV^2$ \cite{sick}. It is normalized as
$\int{d^3 r\rho_C(r) }= 6$. In the second parenthesis, the first term
corresponds to the two
s-wave protons and the second term to the four p-wave protons.
$F_A^{em}(t)$ is obtained by the
Fourier transform of this density, normalized to  $F_A^{em}(0)=1$.

Using the harmonic-oscillator wave functions that give this density in
Eq.~(\ref{15}) we obtain for the
$pA$ non-flip amplitude
\beqn
F^C_0(q) &=&
i\,\int d^2b\, e^{i\vec q\cdot\vec b}\,
\left\{1 -\left[1-\frac{a \,\sigma_{tot} \,(1 -
i\,\rho)}{2\pi (1 + 2 B_{NN}\, a)} \exp\left(-\frac{a \,b^2}{1 + 2
B_{NN}\,a}\right)\right]^4\right.
\nonumber\\ &\times&  \left.
\left[1 - \frac{a \,\sigma_{tot}\,
(1 - i \rho)}{2\pi (1 +2 B_{NN}\,a)}\,
\left(1 -
\frac{2}{3(1 + 2 B_{NN}\,a)} + \frac{2 a\,b^2}{3 (1 +2 B_{NN}\,
a)^2}\right)\right.\right.
\nonumber\\ &\times& \left.\left.
\exp\left(-\frac{a\, b^2}
{1 + 2 B_{NN} \,a}\right)\right]^8\right\}
\label{4.26}
 \eeqn
 \cite{bassel}.
This amplitude provides both imaginary and real part of the amplitude. Thus,
\beq
\rho_{pC}(q)= {\rm Re}F^C_0(q)/{\rm Im}F^C_0(q).
\label{4.27}
\eeq

The hadronic form factor $F_A^h(t)$ is given by
\beq
F_A^h(t)= {\rm Im} F^C_0(q) /{\rm Im} F^C_0(0).
\eeq
In calculating Eq.~(\ref{4.26}) we have taken account of the following
effect: in going from the
charge form factors to the wave-functions we must recognize that the proton
em form factors are
already included in the nuclear em form-factors. The integration in
Eq.~(\ref{4.26}) over the assumed Gaussian form for the amplitude,
with slope $B$ in GeV$^{-2}$, corresponding to a proton radius of
$\sqrt{2 B}$, effectively replaces the nuclear radius squared $R_A^2
\Rightarrow R_A^2 + 2 B$. In
order to avoid including the proton size twice, we correct this for the
means square charge
radius of the proton,
\beq
B_{NN} \Rightarrow
B - {1\over 3}\,\la r^2_{ch}\ra\ .
\label{2.12}
\eeq
so we use the reduced slope $B_{NN} =5.0 \,{\rm GeV}^2$ instead of the
observed hadronic slope.
This is a small effect because the nuclear radius is so much bigger.

The Coulomb phase $\delta_{pA}$ in Eq(\ref{4.40}) and Eq(\ref{4.50}) can be
found
using the result of recent calculations \cite{kt},
which in a good
approximation
takes an especially simple form,
 \beq
\delta_{pA}(q) =
\alpha Z_1Z_2\,
e^{2w}\,\Bigl[2\,E_1(2w)-E_1(w)\Bigr]\ ,
\label{4.34}
\eeq
where $w = {1\over 4} q^2 B_{pA}$ with $B_{pA}$ the slope of the $pA$
diffraction peak.

\begin{figure}[thb]
\centerline{\epsfbox{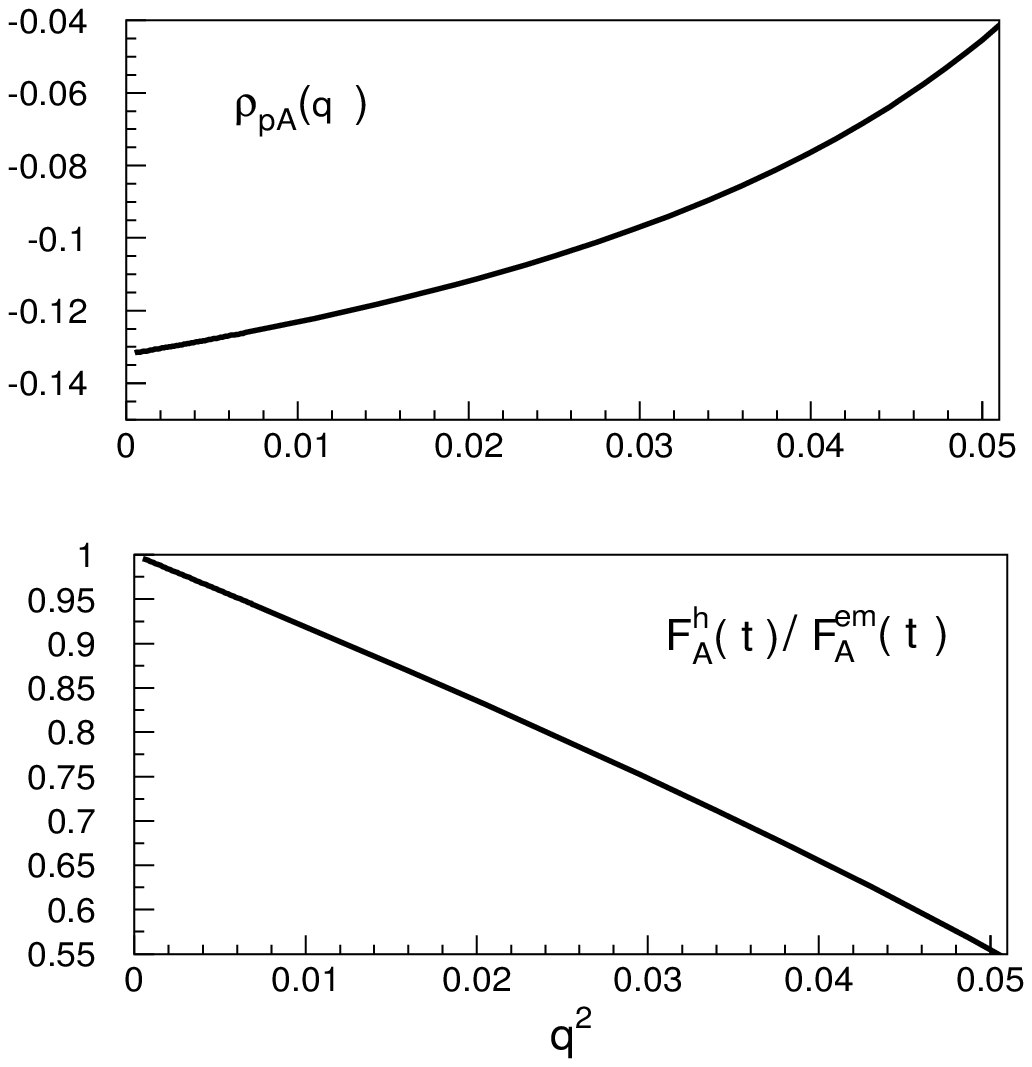}}
\medskip
{\caption[Delta]{\it The upper panel: ratio of real to imaginary parts
of the elastic amplitude for proton-carbon scattering calculated with
Eq.~(\ref{4.26}). The bottom panel: Ratio of the hadronic
 and electromagnetic  form factors as
function of $t=-q^2$.}
\label{rho}}
\end{figure}

Using these results we calculate for $E_p = 22$ GeV, appropriate for the
recent AGS experiment
E950, and show in Fig.~\ref{rho} both
$\rho_{pA}(q^2)$ and $F_A^h(q^2)/F_A^{em}(q^2)$ over the very small
$t$-range. We used
$\sigma_{\rm tot}=39$ mb at this energy and the
parameterization from \cite{pD} for energy dependence of the real
amplitude of elastic proton-deuteron scattering, $\rho_{pD}(s_{pN})=
-0.45 +0.07\,\ln(s_{pN})$.
 Note that $\rho_{pC}(q^2)$ steeply decreases with $|t|$. As usual we employ
the relation $B_{pA}\gg B$ for the proton which allows us to use in
Eq.~(\ref{4.26})
the  $pN$  $\rho$ value at $t=0$. The $t$-dependence of $\rho_{pC}$
shown in Fig.~\ref{rho} is a result of nuclear effects. Likewise, the
hadronic form factor is
seen to drop off much faster than the em form factor. Both of these nuclear
properties will be
seen to have a significant effect on $A_N$, especially for the larger
values of $q^2$ in this
range.

\begin{figure}[tbh]
\centerline{\epsfbox{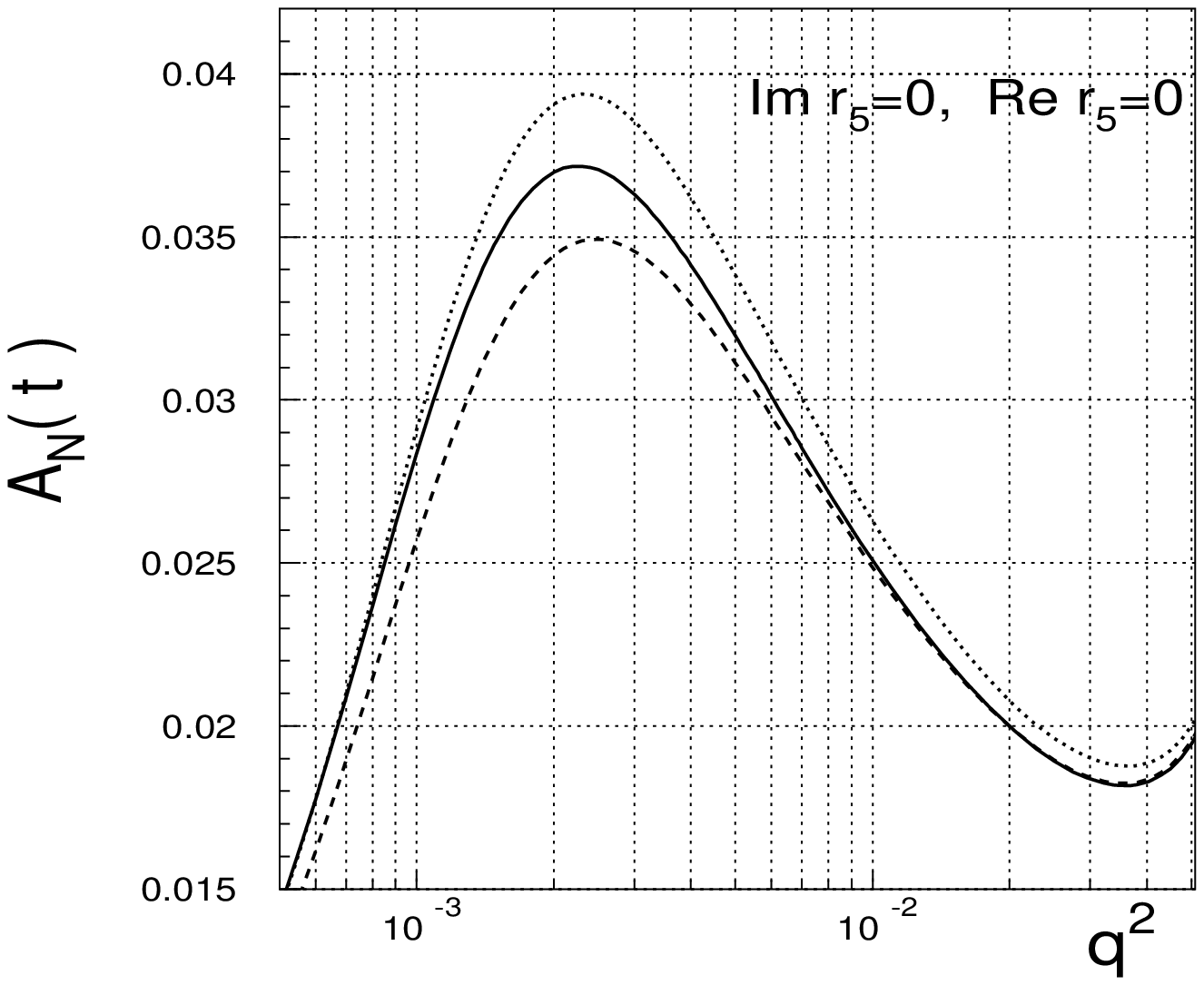}}
\medskip
{\caption[Delta]
{\it Asymmetry in polarized proton scattering on
carbon in the Coulomb-nuclear region of momentum transfer.
All curves are calculated using Eqs.~(\ref{4.40})-(\ref{4.27})
with $|r_5|=0$. The dotted curve corresponds to
$\rho_{pC}=\delta_{pC}=0$. The dashed curve is calculated with
$\rho_{pC}$ shown in Fig.~\ref{rho} and $\delta_{pC}=0$.
The solid curve includes effects of both $\rho_{pC}$ and
$\delta_{pC}=0$.}
\label{an1}}
\end{figure}

Now we are in position to calculate the CNI asymmetry for elastic
proton-nucleus scattering. First we calculate $A_N(t)$ assuming no
hadronic spin-flip and fixing $\rho_{pC}=\delta_{pC}=0$. The result is
depicted by dotted curve in Fig.~\ref{an1}.
 The dashed curve includes $\rho_{pC}$ calculated with Eq.~(\ref{4.27}),
but with $\delta_{pC}=0$.  The solid curve shows the full calculation
including both $\rho_{pC}$ and $\delta_{pC}$ calculated with
Eq~(\ref{4.27}) and Eq~(\ref{4.34}). We keep ${\rm Re}~r_5=0$ in all
calculations.
We see that the asymmetry is quite sensitive to the
values of both $\rho_{pC}$ and $\delta_{pC}$; however they contribute with
opposite
signs at small $t$ and partially cancel.

\begin{figure}[thb]
 \centerline{\epsfbox{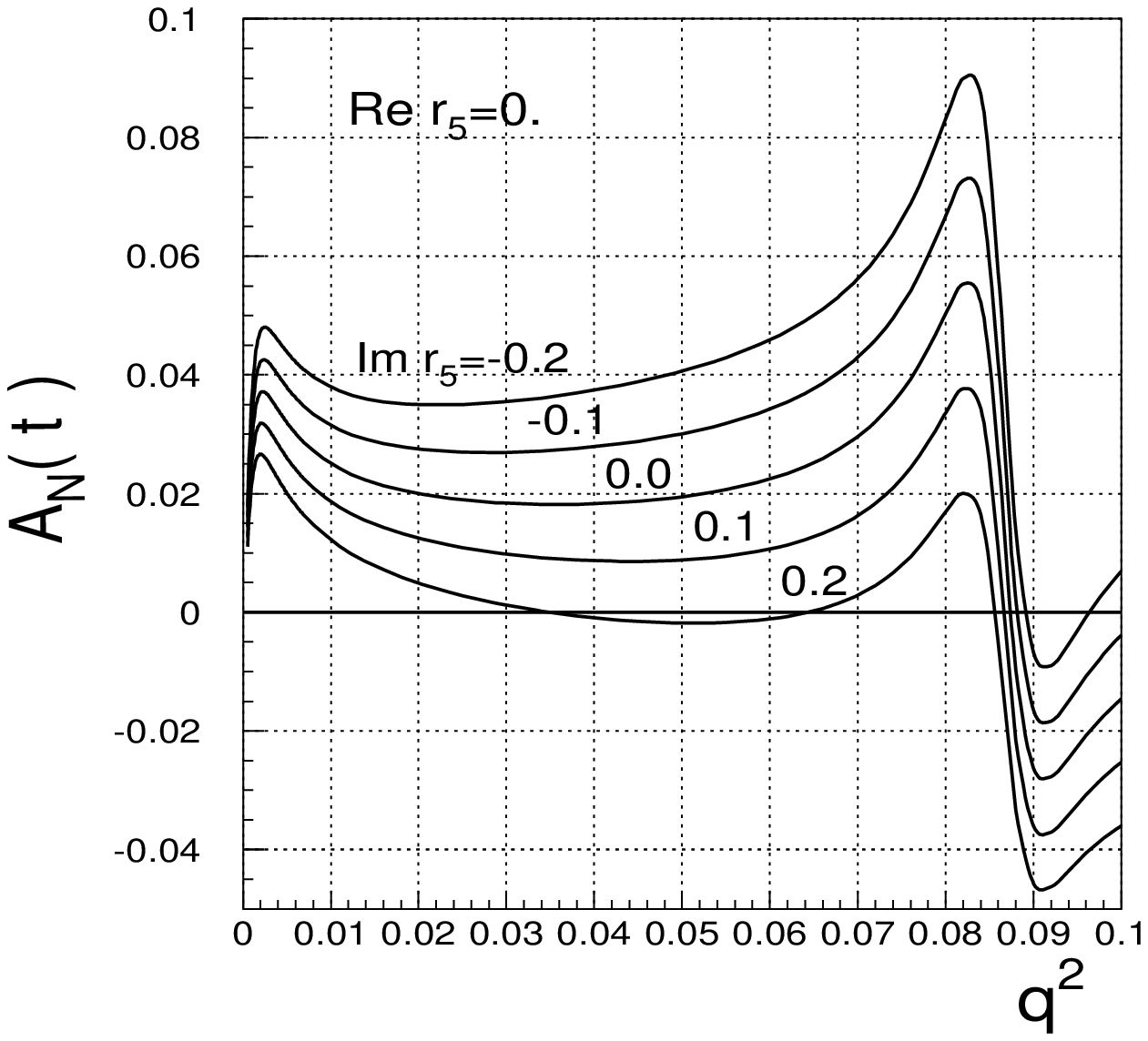}}
\medskip
{\caption[Delta]
{\it The same as in Fig.~\ref{an1}, with ${\rm Re}~r_5=0$,
but ${\rm Im}~r_5=-0.2,\ -0.1,\ 0,\ 0.1,\ 0.2$.
The $\rho_{pC}(q)$ and $\delta_{pC}(q)$ are included as calculated
with Eqs.~(\ref{4.27}) and (\ref{4.34}).}
\label{an2}}
\end{figure}

We next examine the behavior of $A_N(t)$ at larger $q^2$
including the region of the minimum in the hadronic form factor of
$C^{12}$. This is shown in Fig.~\ref{an2} for several reasonable values of
$r_5$.
In spite of smallness of the Coulomb amplitude at such large
$q^2$ the hadronic non-flip amplitude passes through  zero near 0.09
GeV$^2$ and is equal to
the electromagnetic near the zero
position. This is why $A_N(t)$ reaches both a maximum and a minimum in
this region.

Fig.~\ref{an2} also shows that the asymmetry $A_N(t)$ is a sensitive
function of ${\rm Im}~r_5$. This property was suggested in \cite{kz} as a
unique way to measure the spin-flip part of the Pomeron amplitude. This
value usually escapes observation since one does not expect a large phase
shift between the spin-flip and non-flip part of the Pomeron amplitude.
Such an experiment needs a polarized proton beam of known polarization
(like in E950 experiment). On the other hand, if one needs to measure
the polarization of a beam the unknown spin-flip hadronic amplitude can
spoil the CNI method of polarimetry \cite{larry}. As soon as the E950
experiment at BNL provides information about $r_5$, one can use it
for polarimetry at high energies at RHIC.


\noi
{\large\bf 5. Summary}

Proton-nucleus elastic scattering in the CNI region seems to be a better
tool to measure the
spin-flip part of the Pomeron amplitude than $pp$ because the main source
of the spin-flip, the
isovector Reggeons, is excluded (for an isoscalar nucleus like $C^{12}$) or
suppressed by $1/A$.
Therefore, one can perform measurements at rather low energies.

The observation made in this paper that the spin-flip fraction of the
amplitude $\mu_P$ is nearly
$A$-independent is very important for the method since it allows the
application of results of
measurements in $pA$ collisions to the $pp$ case. We have estimated the
corrections to this
statement caused by possible $t$-dependence of $\mu_P$ or by inelastic
shadowing corrections and
found these effects to be small.

We have predicted the asymmetry $A_N(t)$ at small $t$ in the kinematic
region corresponding to
the forthcoming data from the E950 experiment at BNL for proton-carbon
elastic scattering with
various assumptions about $\mu_P$. The real part of the $pC$ elastic
amplitude and the Coulomb
phase have been calculated and incorporated into our predictions.
Comparison of E950 data is
expected to provide the first determination of the spin properties of the
Pomeron.

As soon as the spin-flip part of the Pomeron amplitude is known, one can
use the CNI method as a
reliable polarimeter for polarized proton beams at RHIC.

\end{document}